\documentstyle[psfig]{l-aa}
\input psfig.tex
\def\ref{\bibitem{}}
\begin{document}
\thesaurus{}
\title{Evidence for    
Gamma-ray Bursts Originating Within 11 Mpc}
\author{Y. Chen~~M. Wu~~\and L. M. Song}
\institute {High Energy Astrophysics Lab, Institute of High Energy
 Physics, Chinese Academy of Sciences, Beijing 100039, PRC}
\date{Received ?; accepted ?}
\maketitle
\markboth{}{Y. Chen et al.}
\begin{abstract}
We investigate the number of gamma-ray bursts in two particular strips
of the sky using the data in 3B catalog of Burst and Transient Source 
Experiment (BATSE). One stripe is related to the plane in which   
the intergalactic globular clusters ($R>\,$25kpc) and Galactic satellite 
galaxies (45kpc$\,<R<\,$280kpc) concentrate, the other   
is concerned to nearby galaxies (1Mpc$\,<R<\,$11Mpc).
We find that the density of GRBs in these two strips
is higher than that in other parts of the sky with significance $2.8$ 
and $1.9\sigma$ respectively.
We also compare the peak flux distribution of GRBs in these two stripes with
that in other parts of the sky, and find no difference in the former stripe
but a difference in the latter with a significant level $\alpha =0.05$.
This is consistent with the distance scales of these two planes.  
So it suggests that at least a substantial fraction of GRBs may be related 
to those objects in these two planes and thus originate within 11 Mpc.
\end{abstract}
\section{Introduction}
Gamma-ray bursts (GRBs) have been discovered for more than twenty years,
but their distances are still unknown.
The spatial distribution pattern is a major clue to their origin.
More than one thousand GRBs have been detected by 
 the Burst and Transient Source Experiment (BATSE)
on the Compton Gamma-ray Observatory (CGRO). However, their locations  
are still consistent with large-scale isotropy (Briggs et al. 1996). 
One possible reason is that
GRBs originate at cosmological distances (Prilutski $\&$ Usov 1975; 
Woods $\&$ Loeb 1994; Paczynski $\&$ Xu 1994). However, Many authors insist 
that GRBs may originate from a Galactic halo (Fishman 1979; Jennings 1982;
Brainerd 1992;) or Galactic disk plus halo (Smith $\&$ Lamb
1993). Scharf (1995) have found some evidence for a local
origin of the GRBs in fluence- and number-weighted dipoles measurement.

If GRBs are cosmological, their spatial distribution should
be isotropic. However, the intergalactic globular clusters ($R>\,$25kpc)
and Galactic satellite galaxies (45kpc$\,<R<\,$280kpc), and
 nearby galaxies (1Mpc$\,<R<\,$11Mpc) concentrate towards respective planes. 
Therefore, if a substantial fraction 
of GRBs are related to these objects, 
The spatial distribution of GRBs may deviate slightly from isotropy,
i.e. the density of GRBs may be higher in these planes than 
in other parts of the
sky. Furthermore, since the possible objects in different planes 
have different distance scales, the excess of GRBs should mostly 
exist in the respective distance scale and this may affect the
peak flux distribution of GRBs.

We introduce these two planes in Sect. 2, 
and in Sect. 3 we analyze the
spatial distribution of GRBs in the 3B catalog of the Burst 
and Transient Source Experiment (BATSE). The result shows
that GRBs concentrate towards these two planes. 
Their distributions of peak flux are consistent with the distance scales
of these two planes.
 Then we calculate the probability of coincidence and give
the significant level.
We give some discussion and draw the conclusion which
suggests that at least
a substantial fraction of GRBs may be not cosmological (Sect. 4).

\section{The Magellanic Group plane and the nearby galaxies plane.}

Many objects usually concentrate
towards a certain plane. In this paper we are interested in two of them.
The first plane is related to the nearby (25kpc$\,<R<\,$280kpc) dwarf 
galaxies and intergalactic globular clusters (Majewski 1994; Schmidt 1992,
 1993). There is reasonable possibility that some of them may be tidal
debris of the Magellanic Clouds (Kunkel 1976; Lin 1993).
They seem to concentrate towards the orbital plane 
named the Magellanic Group plane (the MG-plane) (see Fig. 1). 
The normal to this plane points to the direction 
$(l,b)=(169^{o},-23^{o})$ (Kunkel 1976). If GRBs are related to these objects,
they will show some concentration towards the MG-plane.
In another case, if GRBs are related to an extended Galactic halo of
primordial objects, these primordial objects may also concentrate towards 
the MG-plane by the gravitational perturbation (Maoz 1993).
The nearby galaxies (1Mpc$\,<R<\,$11Mpc) seem to concentrate towards another 
plane. We may call it the nearby galaxies plane (the NG-plane). The 
majority of the nearby galaxies 
are situated around this plane whose normal direction points approximately to 
$(l,b)=(47^{o},3^{o})$. Fig. 2 shows the
distribution of these 243 nearby galaxies (Schmidt 1992, 1993).

%fig. 1
\begin{figure}
\psfig{figure=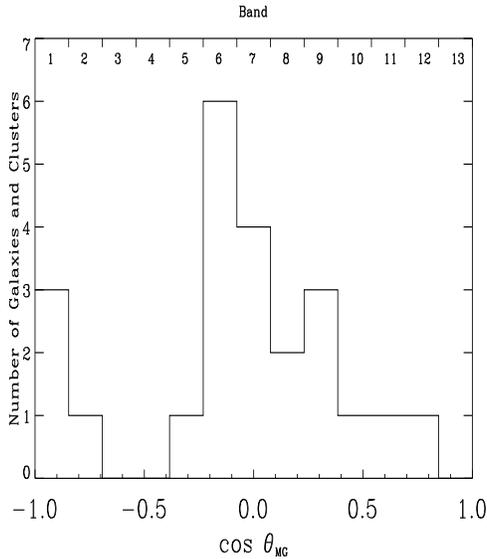,height=7.02truecm,width=6.94truecm,angle=0}
\caption[]{ the spatial distribution of nearby satellite 
galaxies and clusters with 25kpc$\,<R<\,$280kpc. 
The whole sky has been divided into 13 bands along the normal direction
of the MG-plane. 
Each band has an equal area.
$\theta_{MG}$ is the angle between a point and the normal direction of 
the the MG-plane $(l,b)=(169^{o},-23^{o})$.}
\end{figure}

%fig. 2
\begin{figure}
\psfig{figure=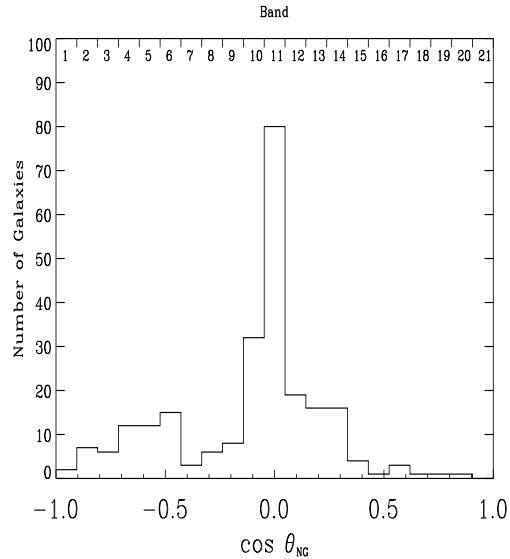,height=7.02truecm,width=6.94truecm,angle=0}
\caption[]{The
distribution of nearby 243 galaxies with a distance 1Mpc$\,<R<\,$11Mpc.
The whole sky has been divided into 21 bands along the normal direction 
of the NG-plane. 
Each band has an equal area.
 $\theta_{NG}$ is the angle between a point and the normal direction of 
the the NG-plane $(l,b)=(47^{o},3^{o})$.}
\end{figure}

If GRBs are related to those objects mentioned above,
they may concentrate towards these two planes.
  
\section{Analysis of the observational data}

The statistics $<\sin^{2}b-1/3>$ is often used to test 
the concentration towards the Galactic plane.
 Maoz (1993) used another similar statistics $<\cos^{2}b_{plane}>$
to determine the concentration towards other particular planes, 
where $b_{plane}$ is the angular distance to the interested plane.
 He found that the value of $<\cos^{2}b_{plane}>$ 
had an increase of 1-2$\sigma$ for the MG-plane by
using BATSE data before March 1992 (241 bursts). 
These statistical methods are powerful to test the quadrupole moment.
They have some advantages. For example, they do not need to
bin the data. But in our case we are only interested
in whether the density of GRBs in these planes is higher than that
in other parts of
the sky. It is not equivalent to the quadrupole moment, 
because, for example, a concentration towards both disk and poles may
not be detected by these quadrupole moment statistics. 
Moreover, if we want to detect the enhancement in more than
one plane, these statistics will be ineffective. 
We can simply divide the whole sky into three parts due to the physical
reasons which will be discussed below.
Two of them (selected area) consist of the interested plane and the
last one (unselected area) does not. We can count the GRBs in each part, 
compare the counts in each selected area with that in 
the unselected area and
calculate the corresponding probability of coincidence.  
According to Fig. 1, it is reasonable to select band 6,7,8 and 9 as
selected area since
they contain about 65$\%$ of the total galaxies and clusters 
with only 30$\%$ of the
sky region. For the NG-plane we select band 10,11 and 12 according to Fig. 2. 
In practice, we let $-0.2<\cos\theta_{MG}<0.4$ for the MG-plane and 
$-0.12<\cos\theta_{NG}<0.12$ for the NG-plane. These two stripes
is our selected area, and we call them Area 1 and Area 2 respectively.
The unselected area is called Area 3 (see Fig. 3). Notice there is 
an overlap between Area 1 and Area 2. Below we will make some test
to investigate the concentration in Area 1 and Area 2 respectively.
The null hypothesis is that the distribution of GRBs is isotropy.  

%fig. 3
\begin{figure*}
\psfig{figure=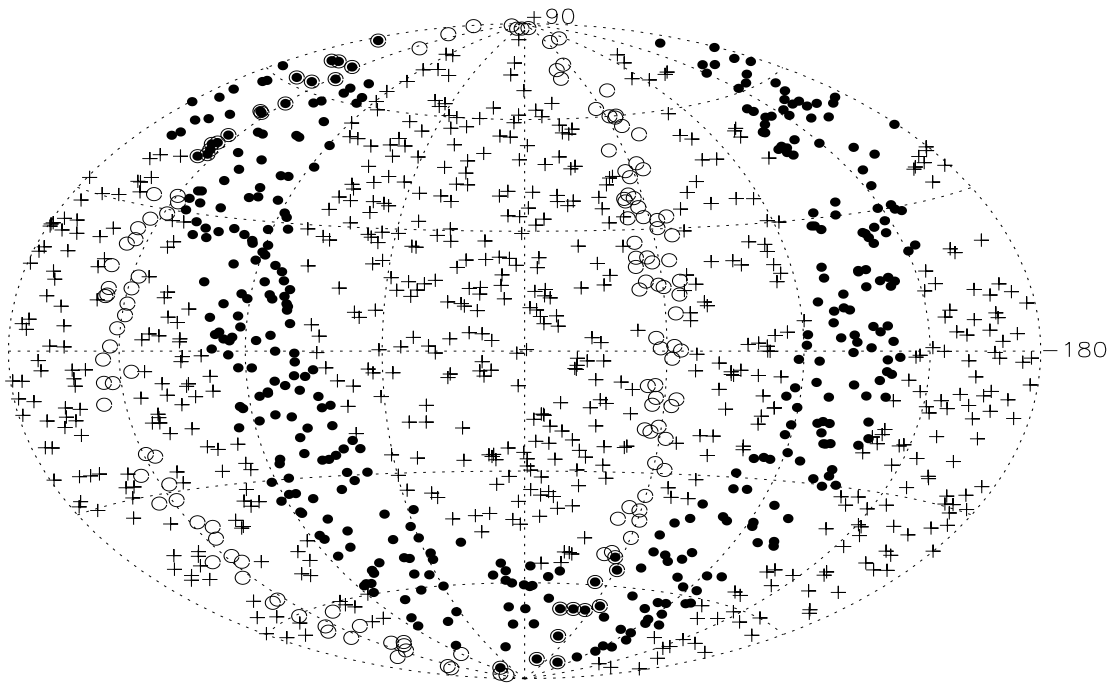,height=10truecm,width=16truecm,angle=0}
\caption[]{The Aitoff projection map of GRBs. The filled circles
represents the
GRBs in Area 1 (MG-plane), the open circles Area 2 (NG-plane) and
the cross Area 3 (unselected area). The 27 filled plus open circles 
represents the overlap. The total number of GRBs is 1122.}
\end{figure*}

\subsection {Density and peak flux distribution in the MG-plane}
Now we compare the counts of GRBs in Area 1 (MG-plane)
with that in Area 3 (unselected area).
The total number of bursts in Area 1 and Area 3 is 
$n=995$ in the 3B catalog. The number of bursts in Area 1 is $m=382$. 
If BATSE's sky exposure is considered here, the normalized equivalent area 
of Area $i$ is 
\begin {equation}
\displaystyle s'_{i}=\int_{s_{i}}t(s)ds / \int_{s_{0}}t(s)ds,
\end{equation}
where $t(s)$ is the exposure time adopted from the first 
BATSE catalog (Fishman et al. 1994), 
$s_{0}$ is the total area of whole sphere and assumed to be 1. 
After sky exposure correction, the areas of Area 1 and Area 3 
are $s_{1}^{'}=0.308$ and $s_{3}^{'}=0.596$ respectively obtained
from Monte Carlo calculation.
We assume the distribution of GRBs is isotropy, thus the number of 
GRBs in Area 1 and Area 3 will follow the binomial distribution.
Therefore, the probability of coincidence is 
\begin {equation}
\displaystyle p=\sum_{i=m}^{n}C^{i}_{n}s_{p}^{i}(1-s_{p})^{(n-i)},
\end{equation}
where $s_{p}=s^{'}_{1}/(s^{'}_{1}+s^{'}_{3})$.
Then we get $p=2.4\times10^{-3}$ (2.8$\sigma$). 
The detail is presented in Tab. 1 and Fig. 4. So we can conclude that 
the density of GRBs in Area 1 (MG-plane) 
is higher than that in Area 3 (unselected area) with a significance
2.8$\sigma$.

%tab. 1
\begin{table*}
\caption[ ]{The number of GRBs in each Area}
\begin{flushleft}
\begin{tabular}{cccccccc}
\hline	   
Plane & $s$  &   $s'$ & $n$ & $m$ & $M$ & $p$ & Significance \\
      &      &        &     &     &     &     & $(\sigma)$   \\
\hline
Area 1 (MG-plane)  & 0.300 & 0.308 &995&382&345.6& $2.4\times10^{-3}$ & 2.8 \\
Area 2 (NG-plane)  & 0.120 & 0.125 &767&154&140.3& $2.7\times10^{-2}$ & 1.9 \\
Area 3 (unselected)& 0.609 & 0.596 & - &613&668.7&        -          &  -   \\
\hline
\end{tabular}
\end{flushleft}
Note: $M$ is the expected number. $M=1122\times s'$.
\end{table*}

Now we further investigate the peak flux distribution of GRBs in
the MG-plane (see the top one of Fig. 5). 
Here we use one-sided Mann-Whitney U-test (Mann et al. 1947) to
test whether the peak flux distribution in Area 1 is fainter than that in 
Area 3. The Mann-Whitney U-test is a widely used nonparametric
test for difference between two independent samples.
We combine the GRBs in Area 1 and Area 3, and array them according to
their peak flux (1024 ms). 
Then we assign ranks to them, starting with 1 for the brightest one and
$n$ for the $n$th. We calculate the sum of the ranks 
in Area 1 $\Sigma R_{1}$ and the significance  
\begin {equation}
\displaystyle t_{s}=(U_{1}-n_{1}n_{3}/2)/\sqrt{n_{1}n_{3}(n_{1}+n_{3}+1)/12},
\end{equation}
where $U_{1}=n_{1}n_{3}+n_{1}(n_{1}+1)/2-\Sigma R_{1}$, and $n_{1}$ and 
$n_{3}$ are the number of GRBs in Area 1 and Area 3 respectively. 
$t_{s}$ is proved to obey the $t$ distribution.
We get $t_{s}=0.58<t_{\alpha=0.05}=1.65$, where $t_{\alpha =0.05}$ is the 
critical value of the $t$ distribution with a significant level 
$\alpha =0.05$.
The result shows that we can not conclude that the peak flux distribution 
in Area 1 is fainter than that in Area 3 (see Tab. 2).
\subsection {Density and peak flux distribution in the NG-plane}

We adopt the similar test as in Sect. 3.1 to the NG-plane. 
Here we only give the results,
for detail see Tab. 1,3 and Fig. 4. The number of GRBs in Area 2
is $m=154$ and $s_{2}^{'}=0.125$, 
Then we get $p=2.7\times10^{-2}$ (1.9$\sigma$)
and conclude that the density of gamma-ray bursts in the NG-plane (Area 2)
is higher than that in unselected area (Area 3) with a significance
1.9$\sigma$ (see Tab. 1).

The peak flux distribution of GRBs in the NG-plane is shown in Fig. 5 (bottom). 
Since the distances of nearby galaxies in the NG-plane are much greater than 
that of the Galactic halo, we can predict the peak flux of them will be less
than that in Area 3. So we use one-sided Mann-Whitney U-test here.
The null hypothesis $H_{0}$ is: The peak flux distribution
in Area 2 is the same as that in Area 3. The alternatives
hypothesis $H_{1}$ is: The GRBs in Area 2 is fainter than that in Area 3.  
We get $t_{s}=1.69>t_{\alpha =0.05}=1.65$, so we conclude the peak flux
of the GRBs in Area 2 is fainter than that in Area 3 with a
significant level $\alpha =0.05$ (see Tab. 3). 

%fig. 4
\begin{figure}
\psfig{figure=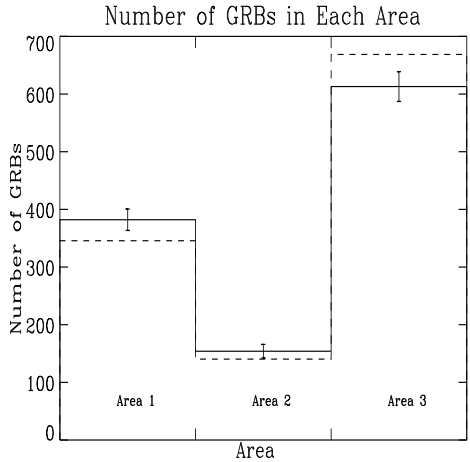,height=7.0truecm,width=7.0truecm,angle=0}
\caption[]{The Number of GRBs in Each Area. The solid lines represent
the number of GRBs in each area and the dashed lines the expected number 
(see Tab. 1). The error bar represents $\pm 1 \sigma$.}
\end{figure}

%fig. 5
\begin{figure}
\psfig{figure=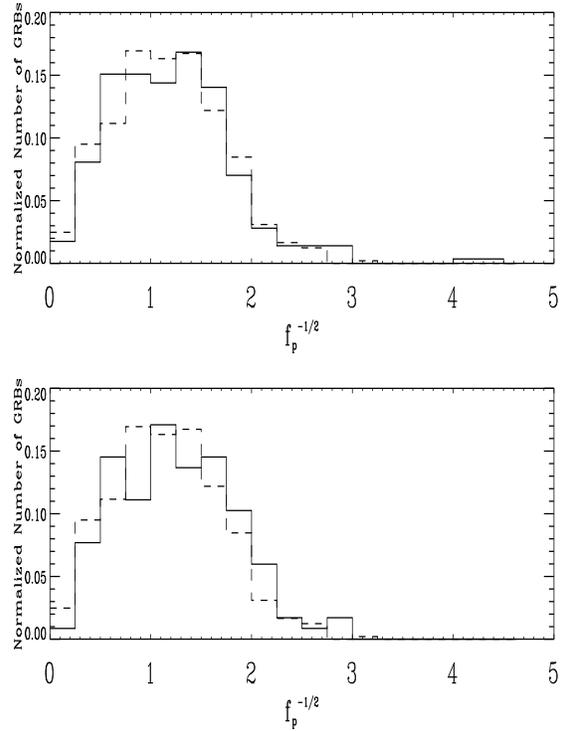,height=10.0truecm,width=8.0truecm,angle=0}
\caption[]{The peak flux distribution in Area 1 (MG-plane, top) and
Area 2 (NG-plane, bottom). $f_{p}$ is the 1024 ms peak flux. 
The solid lines show the peak flux distribution in the selected areas.
For comparison, the dot lines indicate the distribution 
of peak flux in the unselected area. All the number of GRBs is normalized.}
\end{figure}
      
%tab. 2
\begin{table}
\caption[ ]{The one-sided Mann-Whitney U-test of the peak flux distribution
 in the MG-plane  }
\begin{flushleft}
\begin{tabular}{ccccc}
\hline	   
Plane & $i$ & $n_{i}$  & $\Sigma R_{i}$ & $U_{i}$ \\
\hline
Area 1 (MG-plane)  & 1 & 285 & 111462 & 67233  \\
Area 3 (unselected)& 3 & 484 & 184603 & 70707  \\
\hline
\end{tabular}
\end{flushleft}
\begin{flushleft}
\begin{tabular}{c}
$t_{s}=(U_{1}-n_{1}n_{3}/2)/\sqrt{n_{1}n_{3}(n_{1}+n_{3}+1)/12}=0.58$ \\
\hline
\end{tabular}
\end{flushleft}
Note: $U_{1}=n_{1}n_{3}+n_{1}(n_{1}+1)/2-\Sigma R_{1}$, \\
$t_{\alpha=0.05}=1.65$, where $t_{\alpha}$ is the critical value
of the $t$ distribution with a significant level $\alpha$. \\
\end{table}
     
%tab. 3
\begin{table}
\caption[ ]{The one-sided Mann-Whitney U-test of the peak flux distribution 
in the NG-plane }
\begin{flushleft}
\begin{tabular}{ccccc}
\hline
Plane & $i$ & $n_{i}$  & $\Sigma R_{i}$ & $U_{i}$   \\
\hline
Area 2 (NG-plane)  & 2 & 117 & 38073   & 25458  \\
Area 3 (unselected)& 3 & 484 & 142828  & 31170  \\
\hline
\end{tabular}
\end{flushleft}
\begin{flushleft}
\begin{tabular}{c}
$t_{s}=(U_{2}-n_{2}n_{3}/2)/\sqrt{n_{2}n_{3}(n_{2}+n_{3}+1)/12}=1.69$ \\
\hline
\end{tabular}
\end{flushleft}
Note: $t_{\alpha=0.05}=1.65$, where $t_{\alpha}$ is the critical value
of the $t$ distribution with a significant level $\alpha$. \\
\end{table}
\section{Discussion and Conclusion}

Although the spatial distribution of GRBs is 
isotropic in large-scale, we find some  
anisotropic signatures in 3B catalog
of BATSE. The density of GRBs 
in the MG-plane and the NG-plane is higher than that in other 
parts (unselected area) of the sky 
with significance 2.8 and $1.9\sigma$ respectively. The brightness of 
GRBs in the NG-plane is fainter than that in unselected area with a 
significant level $\alpha =0.05$, while no difference is detected
between the peak flux distributions in Area 1 (MG-plane) and in 
Area 3 (unselected area). The distance scale of the MG-plane is 
25kpc$\,<R<\,$280kpc which is similar to that of the Galactic halo.
So it has relatively little effect on the peak flux distribution. 
In the case of the NG-plane, since its distance scale (1Mpc$\,<R<\,$11Mpc)
is much greater than that of the Galactic halo, it has more effect
on the peak flux distribution than the MG-plane and the difference
in peak flux distribution is detected.
The results suggest that GRBs, or at least a substantial fraction of them
may do not originate at cosmological distances. Most GRBs may originate in 
an extended Galactic halo, some faint ones may originate from
nearby galaxies and their extended halos or some unknown
objects which are related to the NG-plane.

\begin{acknowledgements}
This work has been made of data obtained through the Compton
Observatory Science Support Center GOF account, provided by
the NASA-Goddard Space Flight Center. We would like to thank
T. P. Li, X. J. Sun, W. F. Yu, Y. X. Feng, F. J. Lu and H. H. Che 
for helpful discussion. This work is supported by the
National Natural Science Foundation of China under grant 19673010.
\end{acknowledgements}

\end{document}